\documentclass[10pt,conference]{IEEEtran}
\IEEEoverridecommandlockouts
\usepackage{cite}
\usepackage{amsmath,amssymb,amsfonts}
\usepackage{algorithmic}
\usepackage{graphicx}
\usepackage{textcomp}
\usepackage{microtype}
\usepackage{xcolor}
\usepackage{multirow}
\usepackage{svg} % 需使用包

\def\BibTeX{{\rm B\kern-.05em{\sc i\kern-.025em b}\kern-.08em
    T\kern-.1667em\lower.7ex\hbox{E}\kern-.125emX}}

\usepackage[most]{tcolorbox}
\usepackage{hyperref}

\tcbset{
  myfindings/.style={
    colframe=black,          % 边框颜色
    colback=gray!10,         % 正文背景（浅灰）
    colbacktitle=black,      % 标题栏背景（深色）
    coltitle=white,          % 标题文字颜色
    fonttitle=\bfseries,     % 标题字体加粗
    arc=2mm,                 % 圆角半径
    boxrule=1pt,             % 边框粗细
    top=1mm, bottom=1mm,     % 上下内边距
    left=2mm, right=2mm,     % 左右内边距
    title style={enhanced},  % 启用标题栏增强功能
  }
}

\makeatletter
\newcommand{\linebreakand}{%
  \end{@IEEEauthorhalign}
  \hfill\mbox{}\par
  \mbox{}\hfill\begin{@IEEEauthorhalign}
}
\makeatother

\begin{document}

\title{Optimizing Token Consumption in LLMs: A Nano Surge Approach for Code Reasoning Efficiency \thanks{* Corresponding authors}}

\author{
\IEEEauthorblockN{Junwei Hu}
\IEEEauthorblockA{\textit{School of Computer Science and Technology} \\
\textit{Tongji University}\\
Shanghai, China \\
2153393@tongji.edu.cn}
\and
\IEEEauthorblockN{Weicheng Zheng}
\IEEEauthorblockA{\textit{School of Computer Science and Technology} \\
\textit{Tongji University}\\
Shanghai, China \\
2154286@tongji.edu.cn}
\linebreakand
\IEEEauthorblockN{Yihan Liu}
\IEEEauthorblockA{\textit{School of Computer Science and Technology} \\
\textit{Tongji University}\\
Shanghai, China \\
2352755@tongji.edu.cn}
% \linebreakand
% \IEEEauthorblockN{Haoda Zhang}
% \IEEEauthorblockA{\textit{School of Emerging Technology} \\
% \textit{University of Science and Technology of China}\\
% Anhui, China \\
% dd20020810@mail.ustc.edu.cn}
\and
\IEEEauthorblockN{Yan Liu*}
\IEEEauthorblockA{\textit{School of Computer Science and Technology} \\
\textit{Tongji University}\\
Shanghai, China \\
yanliu.sse@tongji.edu.cn}
}

\maketitle

\begin{abstract}
With the increasing adoption of large language models (LLMs) in software engineering, the Chain of Thought (CoT) reasoning paradigm has become an essential approach for automated code repair. However, the explicit multi-step reasoning in CoT leads to substantial increases in token consumption, reducing inference efficiency and raising computational costs, especially for complex code repair tasks. Most prior research has focused on improving the correctness of code repair while largely overlooking the resource efficiency of the reasoning process itself. To address this challenge, this paper proposes three targeted optimization strategies: \textbf{Context Awareness}, \textbf{Responsibility Tuning}, and \textbf{Cost Sensitive}. \textbf{Context Awareness} guides the model to focus on key contextual information, \textbf{Responsibility Tuning} refines the structure of the reasoning process through clearer role and responsibility assignment, and \textbf{Cost Sensitive} incorporates resource-awareness to suppress unnecessary token generation during inference. Experiments across diverse code repair scenarios demonstrate that these methods can significantly reduce token consumption in CoT-based reasoning without compromising repair quality. This work provides novel insights and methodological guidance for enhancing the efficiency of LLM-driven code repair tasks in software engineering.

\end{abstract}


\section{Introduction}
\begin{figure}[htbp]
  \centering
  \includegraphics[width=\columnwidth]{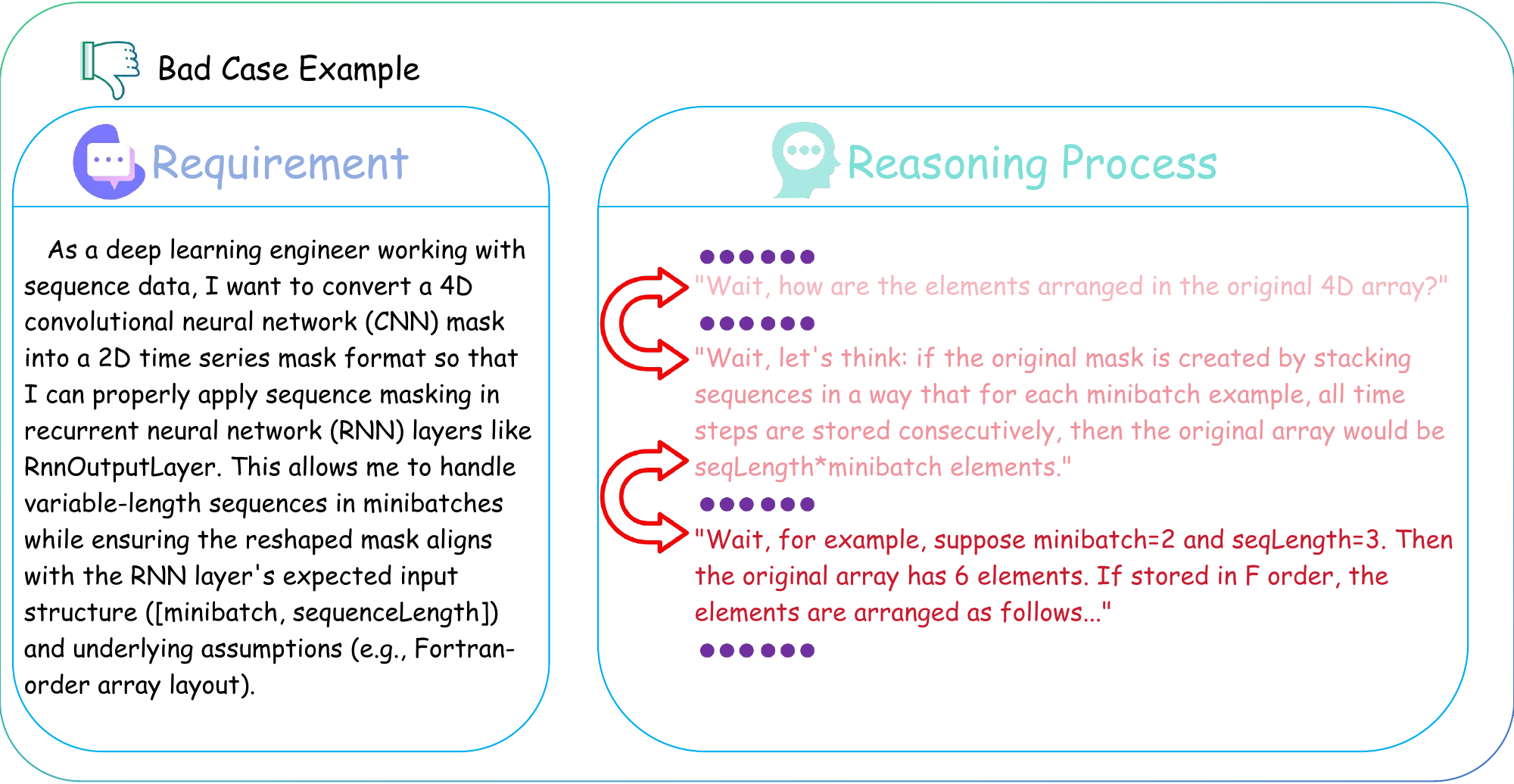}
  \caption{An example of redundant reasoning caused by confusion between C (row-major) and Fortran (column-major) memory layout. The user request involves reshaping a 4D CNN mask into a 2D time series mask compatible with RNNs, which requires correct alignment with Fortran-order expectations. However, the reasoning model repeatedly revisits the same subproblem—how elements are ordered in memory and how reshaping should be done—without converging on a solution. This looped verification, highlighted by repeated statements and backward arrows, illustrates the reasoning inefficiency when the model lacks concrete assumptions about memory layout or reshape semantics. } 
  \label{fig:fortran-c-confusion}
\end{figure}
The rapid proliferation of Large Language Models (LLMs) is driving a profound transformation within the software engineering domain. From code generation and documentation to intricate system design and defect analysis, LLMs exhibit unprecedented potential, fundamentally altering traditional software development practices. Among these advancements, Chain of Thought (CoT)\cite{deepseekai2025deepseekr1incentivizingreasoningcapability} has garnered widespread attention as a strategy designed to enhance the reasoning capabilities of these models. By prompting LLMs to emulate a human-like, step-by-step problem-solving process, CoT significantly improves model accuracy and output quality when tackling complex software engineering tasks, such as generating intricate code, remediating deep-seated errors, and executing multi-step optimizations\cite{cao2023studypromptdesignadvantages}\cite{nikiema2025code}. This progress signifies a crucial shift for LLMs, from mere "pattern matchers" to more sophisticated "intelligent reasoners," thereby unlocking new vistas for automated and intelligent software development\cite{hou2024largelanguagemodelssoftware}.

However, the enhanced capabilities afforded by CoT are accompanied by a significant and often overlooked challenge: a substantial increase in token consumption\cite{kang2025distillingllmagentsmall}. The CoT mechanism inherently relies on the generation of detailed intermediate reasoning steps, which implies that both the length of input prompts and the volume of model-generated content can far exceed those of traditional, more succinct question-answering paradigms\cite{guo2024stopefficientcodegeneration}. This phenomenon, termed "token inflation,"\cite{Dolata_2024} has emerged as a critical bottleneck impeding the broader application of LLMs in the software engineering field. When LLMs are tasked with analyzing large-scale, complex modern software systems, or engaged in programming tasks requiring multiple iterative refinements, the token inflation precipitated by CoT becomes particularly acute\cite{luo2025autol2sautolongshortreasoning}. This not only imposes economic pressure on the cost-effectiveness of model utilization but also adversely impacts practical application efficiency.

In real-world software development applications, cost control and operational efficiency remain paramount considerations\cite{Alarcia_2024}. For all software development teams—ranging from large technology corporations and agile small-to-medium-sized enterprises to budget-constrained startups—the adoption of new technologies transcends mere enhancement of technical capabilities, necessitating a thorough evaluation of cost-effectiveness\cite{guo2024stopefficientcodegeneration}. Currently, a majority of LLM services are accessed via API calls, with charges typically levied based on the volume of tokens processed. While CoT can yield higher-quality outputs, the concomitant substantial increase in token consumption may render its use prohibitively expensive\cite{luo2025autol2sautolongshortreasoning}. Within modern software development pipelines, particularly in contexts such as mobile application development, embedded systems, and other resource-constrained environments, excessive token consumption can lead to considerable economic burdens, compelling many teams to exercise caution or curtail their reliance on CoT methodologies\cite{deepseekai2025deepseekr1incentivizingreasoningcapability}.

Beyond direct economic costs, excessive token consumption engenders other detrimental effects, including prolonged model response times and heightened demands on computational resources. Such issues can impair the agility of software development processes and the immediacy of user experiences, especially in scenarios mandating rapid responses and efficient delivery\cite{guo2024stopefficientcodegeneration}. Consequently, achieving an optimal balance between harnessing the reasoning advantages of CoT and ensuring the economic viability and operational efficiency of LLM applications has surfaced as a significant challenge confronting the software engineering field\cite{yeo2025demystifyinglongchainofthoughtreasoning}.

This study does not aim to propose a systemic solution; rather, it represents a preliminary exploration of this issue. Through the observation and analysis of token consumption phenomena within CoT processes, we aspire to furnish insights for future research, investigating methodologies to effectively curtail token consumption without a significant detriment to reasoning performance. This research endeavors to pave the way for the expanded application of LLMs in software engineering, particularly within contexts that prioritize developmental efficiency and resource optimization.

Driven by widespread concerns regarding the token consumption induced by Chain of Thought (CoT) and its potential ramifications, this study undertakes an exploratory investigation. Our objective is to meticulously observe and comprehend the token dynamics of CoT when processing code of varying complexities within specific experimental settings\cite{1232284}. Furthermore, we aim to identify external intervention strategies capable of effectively mitigating such consumption. To ground our research in realistic and common software development contexts—scenarios where the intricacies of CoT's reasoning pathways and token expenditure may be amplified—we concentrate our observational lens on how Large Language Models (LLMs) handle code exhibiting different intrinsic characteristics. Specifically, the presence and nature of "code smells\cite{10.5555/311424}\cite{1357825}" are employed as a tangible proxy and experimental variable for examining these phenomena. It is crucial to emphasize that our intent is not to present "code smell elimination" as an end in itself, but rather to utilize it as an effective "lens" through which to gain deeper insights into the broader challenge of CoT token optimization\cite{zhang2024comprehensiveevaluationparameterefficientfinetuning}.

Guided by this philosophy, we propose a conceptual methodological framework termed "Token-Aware Coding Flow." Based on this framework, we have conducted a series of preliminary experimental inquiries. These inquiries endeavor to offer initial insights into several key questions:
\begin{itemize}[leftmargin=*,nosep] % nosep 减少列表项间距
	\item What is the specific impact of intrinsic code characteristics (e.g., as studied through a comparative analysis of "clean code" versus "smelly code") on token consumption during CoT reasoning processes?
	\item To what extent does pre-processing code (for instance, by refactoring to eliminate code smells, thereby simplifying the input furnished to CoT) demonstrate potential for reducing token consumption?
	\item Do different types of inherent code complexities (exemplified by various categories of code smells) lead to differentiated patterns in token consumption?
	\item When interacting with LLMs, does explicitly indicating potential code characteristics (such as the type of code smells) within the prompt help guide the CoT process toward more economical token usage?
\end{itemize}

Our preliminary experiments reveal a clear correlation between the intrinsic characteristics of code and token consumption during Chain of Thought (CoT) reasoning. Compared to "clean" code, code with "smells" significantly increases token usage, while refactoring to remove such code smells can substantially reduce the tokens required for subsequent reasoning. Furthermore, different types of code complexity (such as various code smells) affect token consumption differently, with deeper logical or structural issues typically imposing a higher token burden. Explicitly indicating code characteristics in the prompt, combined with prompt engineering strategies, also demonstrates potential for optimizing token efficiency.

Although this research is still in an exploratory phase and its conclusions require further validation, our initial results provide new perspectives on understanding and optimizing token consumption in CoT scenarios.\textbf{Our main contributions of this work are as follows:}
\begin{itemize}
	\item For the first time, we systematically focus on resource (token) consumption in code generation and processing tasks, drawing attention to the economic and efficiency impact of CoT reasoning in LLM-based software engineering.
	\item We propose a novel approach to reduce resource consumption from outside the model, by leveraging code refactoring and prompt engineering techniques. This strategy offers practical and effective ways to optimize token usage without modifying model internals.
	\item We empirically validate that code smells lead to significantly higher token consumption, and demonstrate that external optimization strategies (such as code refactoring and prompt design) can effectively reduce this overhead.
	\item Our work provides empirical evidence and methodological inspiration for developing more cost-effective and efficient LLM-assisted software engineering workflows, contributing to the advancement of resource-aware intelligent development practices.
\end{itemize}

\section{Introduce the New Coding Flow}
\begin{figure}[htbp]
	\centering
	\includegraphics[width=\columnwidth]{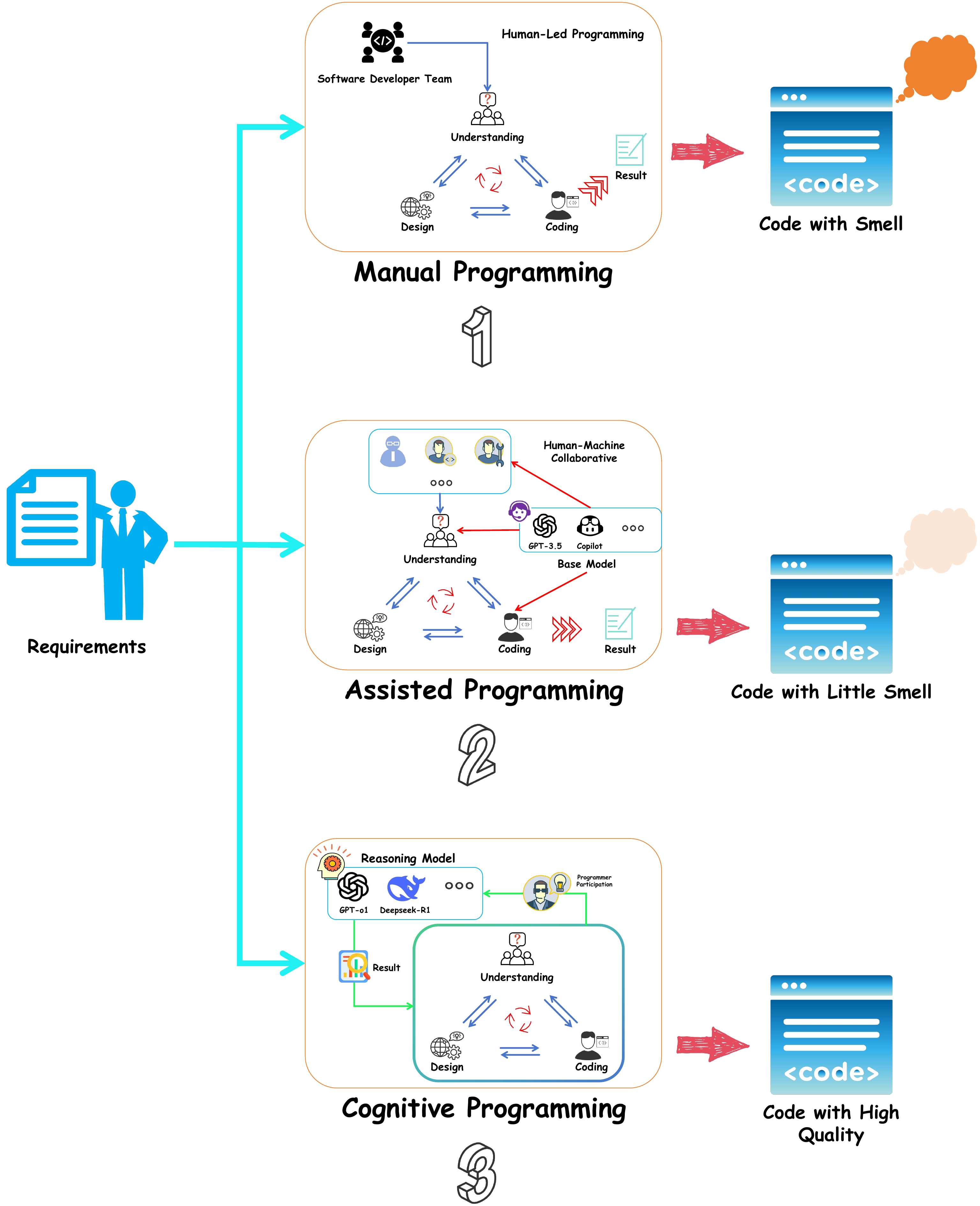}
	\caption{The Ladder of Programming and AI. (1) Traditional coding paradigm: human programmers manually translate requirements into code. (2) Augmented coding paradigm: programmers are assisted by base language models (e.g., GPT-3.5, GitHub Copilot) to generate code, reducing manual effort. (3) Reasoning-driven paradigm: advanced reasoning models (e.g., GPT-o3, DeepSeek-R1) interpret requirements and autonomously generate code by simulating human-like thinking, leading to deeper understanding and improved code quality. }
	\label{coding-evolution}
\end{figure}

The history of software development is characterized by a continuous pursuit of enhanced efficiency, superior quality, and greater automation capabilities. To better contextualize the significance of our proposed "Token-Aware Coding Flow," we first review and analyze the evolution of software development paradigms, particularly highlighting the recent transformations driven by Large Language Models (LLMs). As depicted in Figure~\ref{coding-evolution}, the evolution of software development models can be broadly categorized into the following three core stages\cite{zakharov2025aisoftwareengineeringperceived}\cite{nghiem2024envisioningnextgenerationaicoding}\cite{sun2025surveyneuralcodeintelligence}:

\subsection{Manual Programming}
In the early phase of software development (Stage 1, Fig.~\ref{coding-evolution}), the process relied almost entirely on manual efforts by programmers. Every step—from requirements to design, implementation, testing, and maintenance—depended on their expertise and time. While this approach maximized human creativity, it suffered from low efficiency, limited scalability, difficult knowledge transfer, and a high risk of human errors and inconsistencies.

\subsection{Assisted Programming}
Advancements such as IDEs and code assistance tools ushered in an assisted programming era (Stage 2, Fig.~\ref{coding-evolution}), where programmers leveraged technologies like auto-completion, syntax highlighting, version control, and basic code generators to boost productivity. More recently, LLM-powered tools (e.g., GPT-3.5, GitHub Copilot) have enabled smarter code suggestions, test generation, and documentation drafts. As a result, the programmer’s role shifted towards supervision and validation, marking the start of true human–AI collaboration.

\subsection{Cognitive Programming}
Today, development is increasingly driven by advanced reasoning models (Stage 3, Fig.~\ref{coding-evolution}). Powerful LLMs with Chain of Thought (CoT) capabilities (such as GPT-o1, DeepSeek-R1\cite{deepseekai2025deepseekr1incentivizingreasoningcapability}, GPT-4, DeepSeek Coder) can understand, decompose, and autonomously address complex requirements—planning, generating, and optimizing multi-step code, and even participating in architecture and debugging. Developers now focus on defining goals and making decisions, while models handle much of the cognitive and implementation workload. This paradigm greatly enhances efficiency and problem-solving, potentially transforming software engineering.

However, as discussed in the Introduction, this reasoning-driven paradigm—especially its reliance on CoT—also leads to a sharp rise in token consumption. Optimizing token efficiency is thus essential to fully realize its benefits and ensure sustainable use in diverse development scenarios. This challenge motivates our proposal of the “Token-Aware Coding Flow” and the exploratory work in this study.

\begin{figure}[htbp]
  \centering
  \includegraphics[width=\columnwidth]{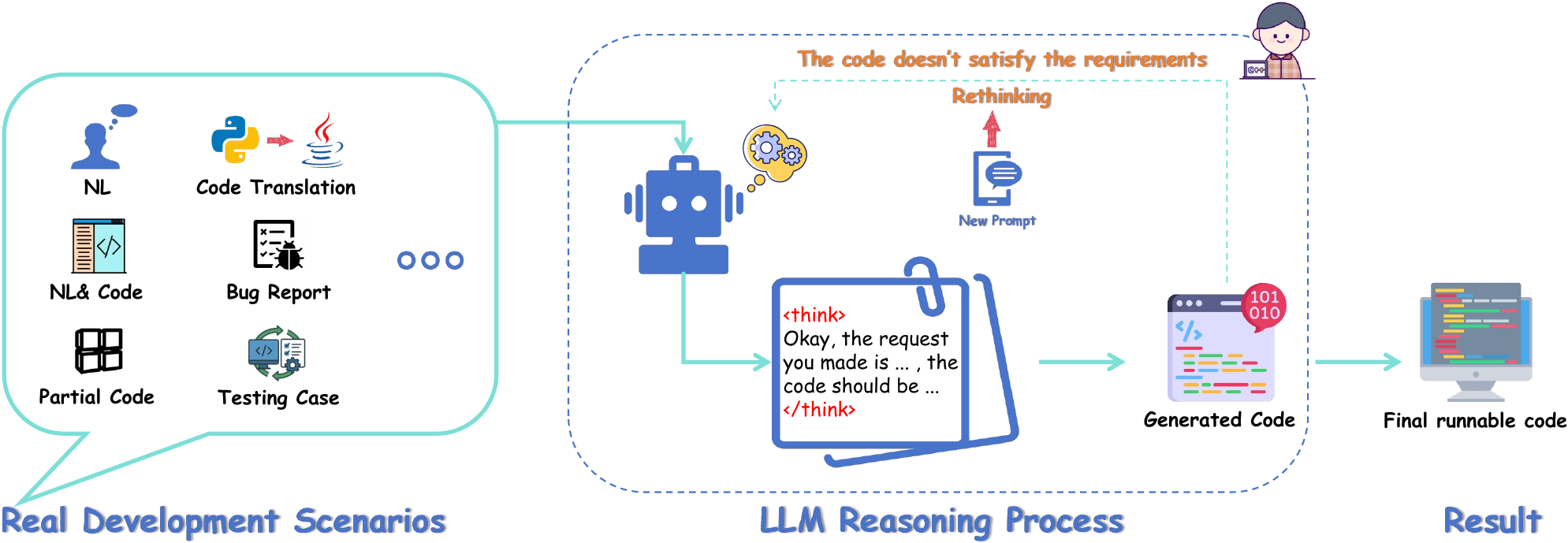}
  \caption{Illustration of the LLM-driven code generation workflow in real-world development scenarios. The process begins with diverse input sources, such as natural language (NL), code translation tasks, bug reports, partial implementations, or test cases. The large language model (LLM) engages in multi-turn reasoning by interpreting the input, thinking step-by-step (e.g., via inner monologue), and generating candidate code. If the generated code fails to meet requirements, the model rethinks the problem by refining prompts and iterating on its reasoning. The process concludes with the production of executable code that satisfies the development goal. } 
  \label{fig:llm-coding-flow}
\end{figure}
\section{Crafting the Future House}
\subsection{Research Questions}
To explore the impact of code smells on the reasoning process of large models, we designed a series of experiments aimed at addressing the following research questions:
\begin{itemize}
  \item \textbf{RQ1:} What are the specific impacts of smelly code on the reasoning process of large models (DeepSeek-R1) compared to clean code?
  \item \textbf{RQ2:} How does code refactoring (eliminating code smells) impact token consumption during model inference?
  \item \textbf{RQ3:} Are there significant differences in token consumption caused by different types of code smells?
  \item \textbf{RQ4:} Does explicitly indicating the type of code smells in the model prompt help reduce token consumption during inference?
  \item \textbf{RQ5:} In addition to code refactoring, what other effective prompt engineering strategies can further reduce token consumption when reasoning with smelly code?

\end{itemize}
Through these questions, we aim to gain a deeper understanding of the impact of smelly code on large model inference and explore the potential improvements in reasoning efficiency with different optimization strategies\cite{zhang2024datapreparationdeeplearning}. First, RQ1 primarily focuses on whether smelly code causes large models to consume more tokens during inference, which is crucial for evaluating the model's ability to handle non-standard code. RQ2 investigates whether refactoring smelly code improves the model's reasoning efficiency, particularly in terms of reducing token consumption while maintaining functional consistency\cite{guo2024stopefficientcodegeneration}\cite{cordeiro2024empiricalstudycoderefactoring}. RQ3 focuses on the impact of different types of code smells on model reasoning efficiency, helping us understand which types of code smells have a more significant effect on token consumption\cite{cordeiro2024empiricalstudycoderefactoring}. RQ4 examines whether explicitly pointing out the type of code smells in the prompt provides the model with more precise contextual information, thereby reducing redundant reasoning steps and lowering token consumption\cite{cordeiro2024empiricalstudycoderefactoring}. Finally, RQ5 explores whether other effective prompt engineering strategies, such as context awareness, role constraints, and cost-sensitive strategies, can further optimize reasoning efficiency beyond code refactoring. By addressing these questions, we aim to both validate the specific impact of code smells on large model reasoning performance and provide experimental evidence for improving inference efficiency through refactoring and optimized prompt design.

\subsection{Tasks}
In this study, we designed a series of evaluation tasks to systematically explore the impact of code smells on the reasoning process of large models \cite{xie2023uncertaintyawaremoleculardynamicsbayesian} and to test the effectiveness of different optimization strategies. The following are the specific tasks set for each research question.

\subsubsection{Observational Experiment}
\textbf{Datasets:}
We used the CodeXGLUE Java (Text-To-Code) dataset, containing 300 smelly code samples and 300 clean code samples, to analyze differences in token and time consumption when DeepSeek-R1 processes different types of code.

\textbf{Metrics:}
We recorded the total number of tokens and inference time for each code sample. Token consumption rate was calculated to evaluate the model's efficiency in handling different code types. The generated code was further assessed for functional correctness and readability.

\subsubsection{Refactoring and Optimization Experiment}
\textbf{Datasets:}
Smelly code samples were refactored using DeepSeek-R1 to generate refactored code, which was then compared with the original smelly code and clean code to assess inference efficiency.

\textbf{Metrics:}
We tracked token consumption at each processing step and calculated Halstead and cyclomatic complexity. Token consumption per unit of complexity and per line of code was analyzed. Functional consistency was ensured using metrics such as CodeBLEU.

\subsubsection{Code Smell Type Comparison Experiment}
\textbf{Datasets:}
We selected 10 types of code smells from the smelly code dataset, with 50 samples per type (500 samples in total), to analyze the impact of different smell types on token consumption.

\textbf{Metrics:}
Token consumption during inference was compared across different smell types, with additional analysis of design, structural, and naming smells. The generated code was also evaluated for functional consistency.

\subsubsection{Explicit Smell Type Prompt Experiment}
\textbf{Datasets:}
Based on the smelly code dataset, we compared token consumption when the prompt explicitly indicated the code smell type versus when it did not.

\textbf{Metrics:}
Token and time consumption were recorded at each experimental step. Functional consistency of the output was evaluated using metrics such as CodeBLEU. The impact of explicitly indicating the code smell type on inference efficiency and code quality was also analyzed.

\subsubsection{Prompt Optimization Strategy Experiment}
\textbf{Datasets:}
Various prompt optimization strategies—such as context enhancement, responsibility tuning, and cost-sensitive prompting—were tested on the smelly code dataset.

\textbf{Metrics:}
Token consumption and code complexity under each strategy were recorded. Token consumption per unit complexity and per line of code was compared. Each strategy was evaluated based on functional consistency and code generation quality.

\subsection{Implementation}
This study’s experimental implementation is built upon existing LLMs and open‑source code‑refactoring tools, with all tasks orchestrated via API calls to guarantee reproducibility and efficiency\cite{hasan2025opensourceaipoweredoptimizationscalene}. Specifically, we employ DeepSeek‑R1 as the inference engine and integrate Tree‑Sitter for code‑smell detection and automated refactoring. The implementation proceeds as follows(Fig.~\ref{fig:code-generation-flow}):

\begin{figure}[htbp]
  \centering
  \includegraphics[width=\columnwidth]{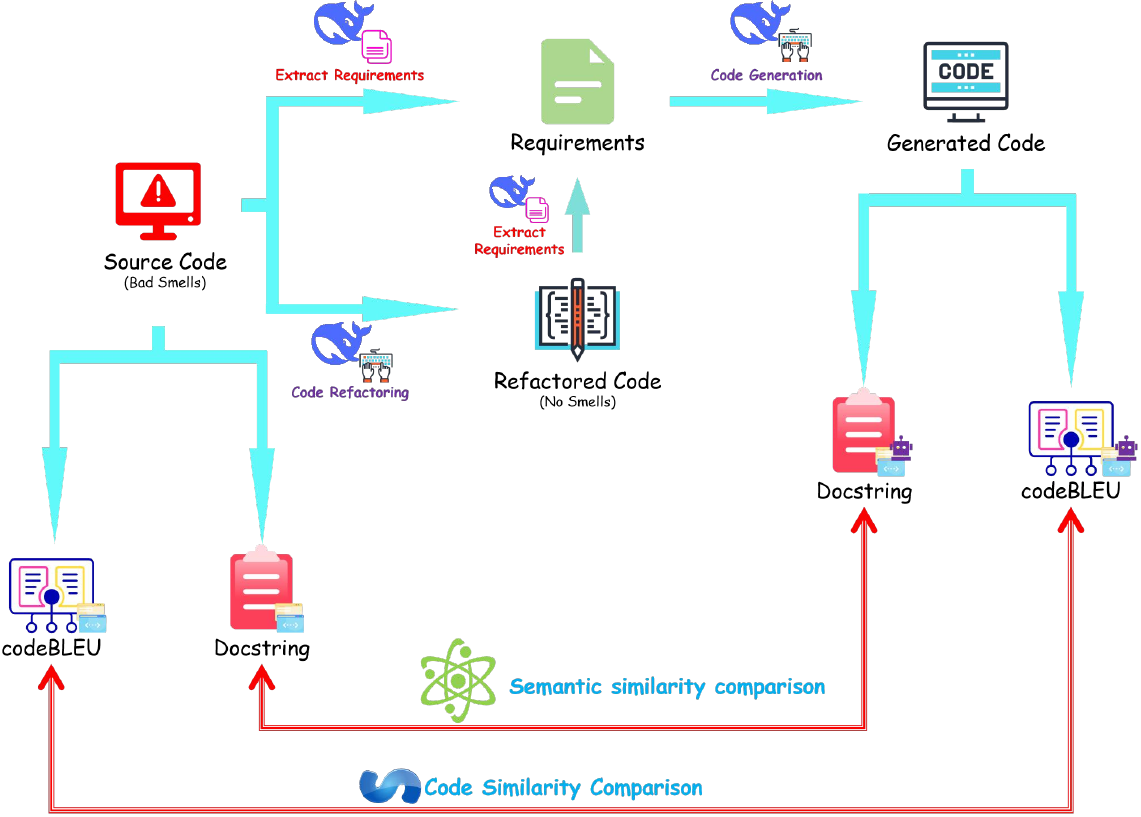}
  \caption{Experiment Design Process. The diagram shows the steps involved in extracting requirements, generating code, refactoring code, and comparing semantic and code similarities. The flow demonstrates the interaction between different components, including code refactoring and requirement extraction, with the evaluation steps represented through similarity comparison.}
  \label{fig:code-generation-flow}
\end{figure}

\begin{itemize}
	\item \textbf{Data Processing and Smell Annotation:} Java code was extracted from the Text‑Code subset of CodeXGLUE and parsed using Tree‑Sitter to annotate ten common code smell categories. This enabled the construction of parallel “smelly” and “clean” datasets. For each sample, we recorded token consumption and inference latency during both code repair and refactoring tasks using the DeepSeek‑R1 API.
	
	\item \textbf{Prompt Engineering and Optimization:} To mitigate token overhead in CoT-based reasoning, we implemented and evaluated three prompt-level strategies: \textbf{Context Awareness}, \textbf{Responsibility Tuning}, and \textbf{Cost Sensitive}. Each strategy was applied by modifying the model's input prompt, and for each experiment, we logged token usage, inference time, and code repair quality.
	
	\item \textbf{Automation and Analysis:} All API interactions were automated via Python scripts, ensuring reproducibility and efficiency. Outputs—including token counts, timing, and repaired code—were collected and analyzed using custom scripts for data aggregation, statistical analysis, and visualization, enabling comprehensive comparison of different strategies.
\end{itemize}

\section{Experimental Study: Triggering the nano surge}
\begin{figure}[htbp] 
	\centering 
	\includegraphics[width=\columnwidth]{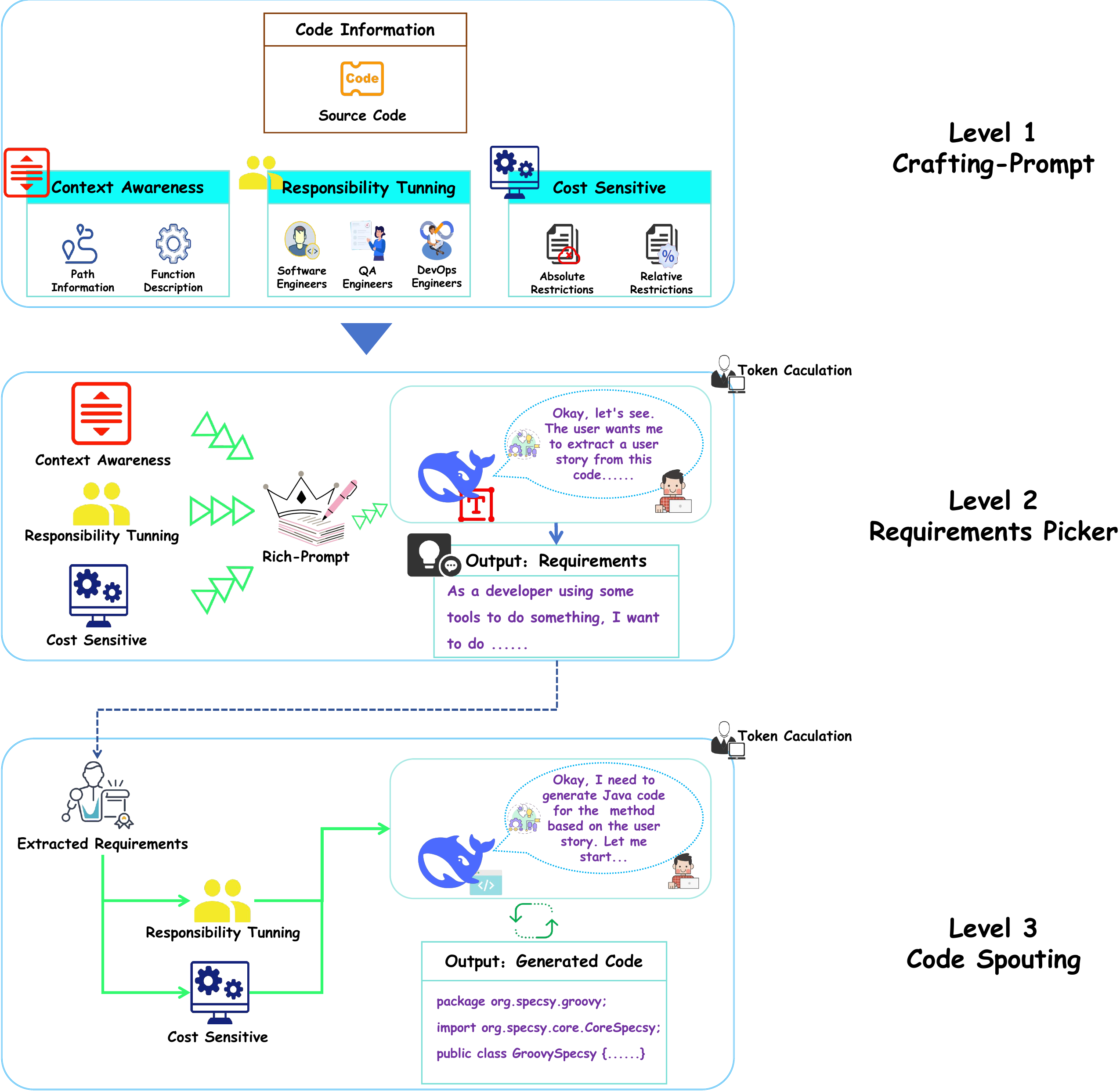} 
	\caption{Nano Surge on Token Consumption} 
	\label{fig:nano-surge} 
\end{figure}

\subsection{RQ1: Smelly Code vs. Clean Code}
\paragraph{Experimental Design.}
We randomly sampled 30 instances from each of the ten individual code‑smell datasets and aggregated them into a single file, smelly\_code.jsonl. We then randomly selected 300 examples from the clean‑code dataset, producing clean\_code.jsonl. Because the two collections differ substantially in content and structure, we employ the \emph{Time‑Scaled Token Consumption} metric to evaluate token expenditure during inference. Using the DeepSeek‑R1 model, each sample is assessed across six dimensions—functional correctness, readability, robustness, maintainability, extensibility, and security—while we record the inference procedure, evaluation outcomes, and end‑to‑end execution time. From these logs, we compute the total tokens consumed during both the reasoning and output‑generation phases.

\begin{figure}[htbp]
	\centering
	\includegraphics[width=0.7\columnwidth]{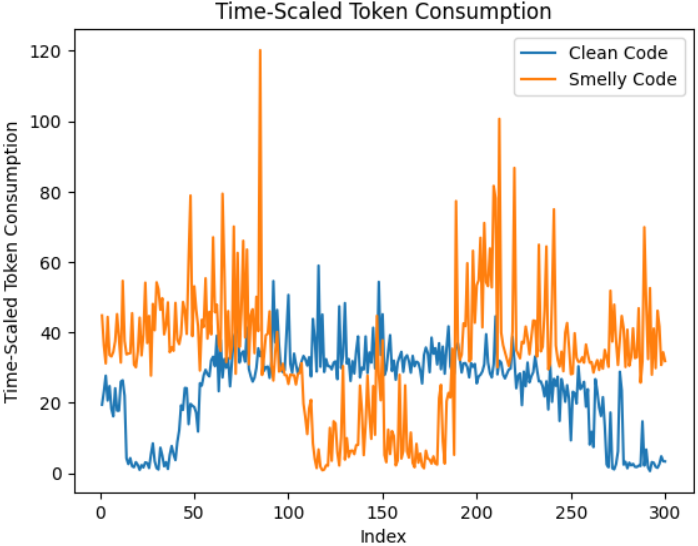}  % Adjusted the width to reduce space
	\caption{Comparison of time-scaled token consumption between clean code and smelly code.}
	\label{fig:time-scaled token consumption}
\end{figure}

\paragraph{Analysis.}
In our analysis, we first compare token consumption per unit time between clean code and smelly code. Directly contrasting absolute token counts would misrepresent efficiency, since the code samples vary greatly in complexity and structure. By normalizing to tokens per unit time, we obtain a more precise measure of model throughput across code types. Statistical evaluation reveals that smelly code consistently demands higher token expenditure(Fig.~\ref{fig:time-scaled token consumption})(Tab.~\ref{tab:token_stats}). This elevated consumption stems from the increased complexity of smelly code: the model must undertake additional reasoning steps and verifications to resolve intricate structural and logical relationships. Conversely, clean code—with its clearer organization—allows the model to complete inference with fewer redundant operations.

\begin{table}[htbp]
	\caption{Descriptive Statistics of Time-Scaled Token Consumption}
	\begin{center}
		\begin{tabular}{p{3cm}|c|c}  % Adjusted column width
			\hline
			\textbf{Statistic} & \textbf{Clean Code} & \textbf{Smelly Code} \\
			\hline
			Count           & 300     & 300     \\
			Mean            & 24.44   & \textbf{33.20}   \\
			Std. Dev.       & 12.35   & \textbf{18.63}   \\
			Min             & 0.61    & \textbf{0.83}    \\
			25\% Quartile   & 18.55   & \textbf{25.30}   \\
			Median (50\%)   & 28.13   & \textbf{33.30}   \\
			75\% Quartile   & 31.68   & \textbf{41.90}   \\
			Max             & 58.97   & \textbf{120.16}  \\
			\hline
		\end{tabular}
		\label{tab:token_stats}
	\end{center}
\end{table}

\begin{tcolorbox}[myfindings,title=Findings\,--\,Smelly Code vs Clean Code]
  The results show that smelly code leads to significantly higher token consumption during inference than clean code. This is mainly attributed to structural complexity such as redundant logic and poor naming, which forces the model to perform repeated validations. These findings suggest that code smells not only harm maintainability but also increase the inference cost of large models, highlighting the importance of code quality for efficient model reasoning.
\end{tcolorbox}
  
\subsection{RQ2: Impact of Code Refactoring on Token Consumption under Functional Consistency Constraint}

\paragraph{Experimental Setup.}
To address RQ2, we examined how refactoring to remove code smells impacts token consumption during model inference. Using DeepSeek‑R1, we first established baseline token usage on both smelly\_code.jsonl and clean\_code.jsonl. We then applied automated refactoring to the smelly code (removing common smells like long methods and duplicated logic) to produce a refactored code set (rf\_code), and measured token usage on these samples. Functional consistency was verified using CodeBLEU and related metrics.

\begin{figure}[htbp]
  \centering
  \includegraphics[width=\columnwidth]{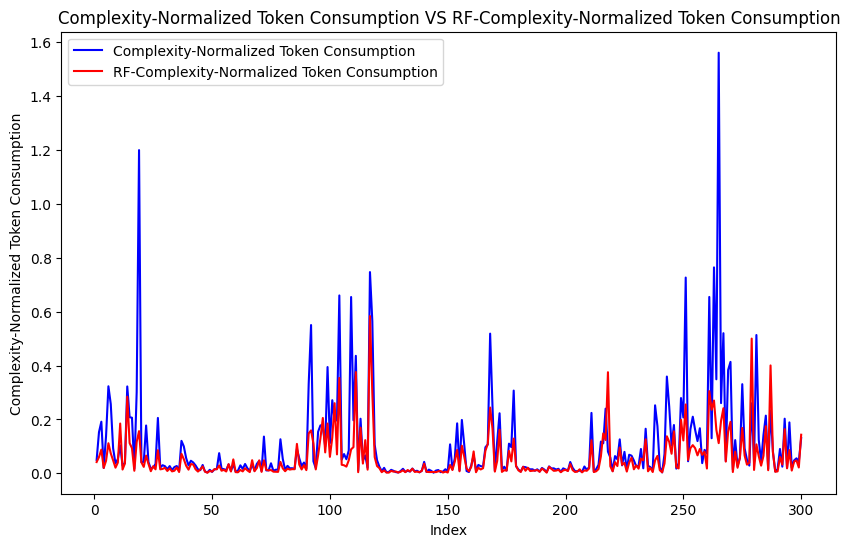}
  \caption{Comparison of complexity-normalized token consumption before and after code refactoring.} 
  \label{fig:token_per_complexity_plot}
\end{figure}

\paragraph{Analysis.}
For each run, we tracked inference duration and token consumption, and normalized by time. Results show that refactored code leads to more efficient inference—maintaining functionality (Tab.~\ref{tab:code_similarity_stats}), but requiring substantially fewer tokens (Fig.~\ref{fig:token_per_complexity_plot}, Tab.~\ref{tab:token_per_complexity}). Refactoring complex smells, especially long functions and duplicate code, significantly reduced the number of reasoning steps and validations the model performed.

\begin{table}[htbp]
  \caption{Descriptive Statistics of Token per Complexity}
  \begin{center}
  \begin{tabular}{p{3.2cm}|c|c}
  \hline
  \textbf{Statistic} & \textbf{Original Code} & \textbf{Refactored Code} \\
  \hline
  Count             & 300     & 300     \\
  Mean              & \textbf{0.1015}  & 0.0576  \\
  Std. Dev.         & \textbf{0.1732}  & 0.0816  \\
  Min               & 0.0020  & 0.0020  \\
  25\% Quartile     & \textbf{0.0144}  & 0.0094  \\
  Median (50\%)     & \textbf{0.0334}  & 0.0239  \\
  75\% Quartile     & \textbf{0.1180}  & 0.0723  \\
  Max               & \textbf{1.5613}  & 0.5836  \\
  \hline
  \end{tabular}
  \label{tab:token_per_complexity}
  \end{center}
  \end{table}
  
  \begin{table}[htbp]
    \caption{Summary of CodeBLEU and Docstring Similarity Statistics (Mean Values Only)}
    \centering
    \begin{tabular}{l|c|c}
      \hline
      \textbf{Category} & \textbf{Metric} & \textbf{Mean} \\
      \hline
      \multirow{2}{*}{\textbf{CodeBLEU}} 
        & code \& rf code  & \textbf{0.5224} \\
        & rf code \& rf gc code  & \textbf{0.1803} \\
      \hline
      \multirow{3}{*}{\textbf{Docstring}} 
        & code \& rf code   & \textbf{0.8604} \\
        & code \& rf gc code   & \textbf{0.7356} \\
        & rf code \& rf gc code  & \textbf{0.7364} \\
      \hline
    \end{tabular}
    \label{tab:code_similarity_stats}
  \end{table}

\begin{tcolorbox}[myfindings,title=Findings\,--\,Refactoring Impact]
  Code refactoring yields a substantial reduction in token consumption during inference while preserving functional correctness. Specifically, the refactored code consumes approximately 50\% fewer tokens compared to the original smelly code. By removing redundant structures and simplifying complex fragments, refactoring not only enhances code quality but also markedly improves inference efficiency. These results suggest that systematic elimination of code smells is a critical strategy for reducing computational overhead in large‑model reasoning.
\end{tcolorbox}

\subsection{RQ3: Impact of Different Types of Code Smells on Token Consumption}

\paragraph{Experimental Design / Setup}
To answer RQ3, we designed an experiment to investigate how different types of code smells affect token consumption in the DeepSeek‑R1 inference model. We extracted from smelly\_code.jsonl ten common smell categories—complicated\_regex\_expression, parameter\_list\_too\_long, binary\_operator\_in\_name, complicated\_boolean\_expression, etc.—with 30 code snippets per category (300 total). We ran DeepSeek‑R1 inference separately on each group, recording the tokens consumed for each snippet. By computing the tokens consumed per unit time for each smell type, we identified which categories impose the greatest inference burden.

\begin{figure}[htbp] 
  \centering 
  \includegraphics[width=0.65\columnwidth]{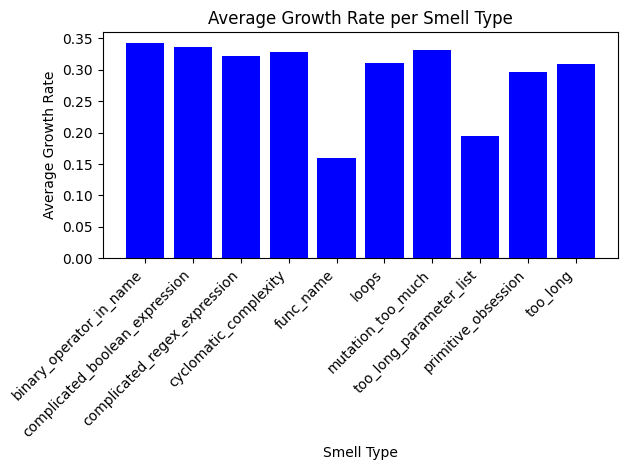} 
  \caption{ Average token growth rate per code smell type.}
  \label{fig:growth_by_smell} 
\end{figure}

\paragraph{Analysis}
The first graph (Fig.~\ref{fig:growth_by_smell}) and table (Tab.~\ref{tab:smell_growth_rate}) present the average growth rate of token consumption for different code smell types. Most smells, such as binary operator in name, complicated boolean expression, and complicated regex expression, cause a notable increase in token usage, likely due to their complex logic requiring deeper model reasoning~\cite{khojah2024impactpromptprogrammingfunctionlevel}. In contrast, smells like func name and too long parameter list have a lower impact, indicating they are less complex and lead to only minor increases in token consumption.

\begin{figure}[htbp] 
  \centering 
  \includegraphics[width=0.65\columnwidth]{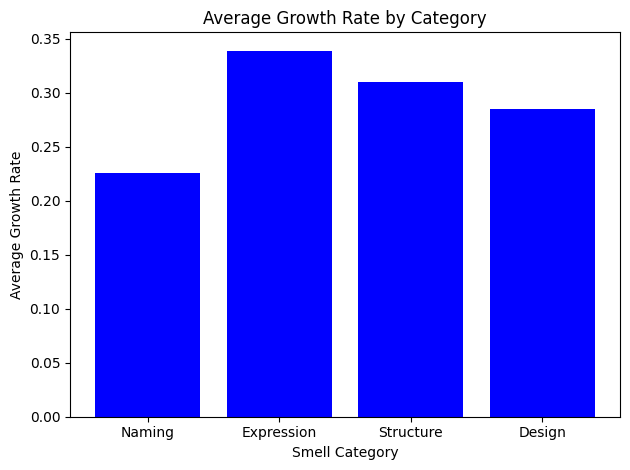} 
  \caption{ Average token growth rate by code smell category.} 
  \label{fig:growth_by_category} 
\end{figure}

The second graph (Fig.~\ref{fig:growth_by_category}) groups code smells into Naming, Expression, Structure, and Design. Expression-related smells show the highest token consumption growth, followed by Structure. These categories often involve intricate logic or convoluted code structures, demanding more tokens for the model to process. By comparison, Naming and Design smells have a smaller effect on token growth, as they are less likely to complicate the reasoning process.

\begin{table}[htbp]
  \caption{Average Token Growth Rate by Smell Type}
  \begin{center}
  \begin{tabular}{p{4.5cm}|c}
  \hline
  \textbf{Smell Type} & \textbf{Avg. Growth Rate} \\
  \hline
  binary\_operator\_in\_name         & 0.3424 \\
  complicated\_boolean\_expression  & 0.3361 \\
  complicated\_regex\_expression    & 0.3221 \\
  cyclomatic\_complexity            & 0.3285 \\
  func\_name                        & 0.1592 \\
  loops                             & 0.3111 \\
  mutation\_too\_much               & 0.3321 \\
  too\_long\_parameter\_list        & 0.1942 \\
  primitive\_obsession              & 0.2963 \\
  too\_long                         & 0.3089 \\
  \hline
  \end{tabular}
  \label{tab:smell_growth_rate}
  \end{center}
  \end{table}

  \begin{table}[htbp]
    \caption{Comparison of Code Equivalence Before and After Refactoring (Mean Values Only)}
    \centering
    \begin{tabular}{l|c|c}
      \hline
      \textbf{Category} & \textbf{Metric} & \textbf{Mean} \\
      \hline
      \multirow{1}{*}{\textbf{Source Code}} 
        & code \& gc code  & 0.1659 \\
      \hline
      \multirow{2}{*}{\textbf{Refactored Code}} 
        & code \& rf code  & 0.5208 \\
        & rf code \& rf gc code  & 0.1773 \\
      \hline
      \multirow{1}{*}{\textbf{Source Code Similarity}} 
        & code \& gc code  & 0.7470 \\
      \hline
      \multirow{3}{*}{\textbf{Refactored Code Similarity}} 
        & code \& rf code  & 0.8610 \\
        & code \& rf gc code  & 0.7295 \\
        & rf code \& rf gc code  & 0.7340 \\
      \hline
    \end{tabular}
    \label{tab:code_equivalence_comparison}
  \end{table}
  
  \begin{tcolorbox}[myfindings,title=Findings\,--\,Smell Type Impact]
    Our study, conducted with a 70\% code functionality similarity threshold(Tab.~\ref{tab:code_equivalence_comparison}), reveals significant variance in token consumption across different code smell categories. Complex code smells—namely complicated\_regex\_expression, parameter\_list\_too\_long, and complicated\_boolean\_expression—incur the highest token costs due to increased code complexity and the need for additional reasoning steps. Simpler smells, such as binary\_operator\_in\_name and func\_name, result in comparatively minor token overhead. These findings suggest that prioritizing the removal of design and structural smells can maximize inference efficiency and substantially reduce computational overhead.
  \end{tcolorbox}

\subsection{RQ4: Impact of Explicitly Indicating Code Smell Types in Prompts on Token Consumption\cite{liu2024promptlearningmultilabelcode}}

\paragraph{Experimental Design / Setup}
To address RQ4, we designed an experiment to investigate whether explicitly annotating code smell types in the model prompt effectively reduces token consumption during inference. We compared two conditions on the same smelly\_code.jsonl dataset: (1) \emph{No annotation}, using the raw code as input; and (2) \emph{Explicit annotation}, where the prompt was augmented with statements such as “the code contains complex regular expressions” or “the code has an excessively long parameter list.” For each condition, we invoked DeepSeek‑R1 to perform inference and recorded both total token usage and inference time. We then computed token consumption per unit time to facilitate a direct comparison between annotated and unannotated prompts.

\begin{figure}[htbp] 
  \centering 
  \includegraphics[width=\columnwidth]{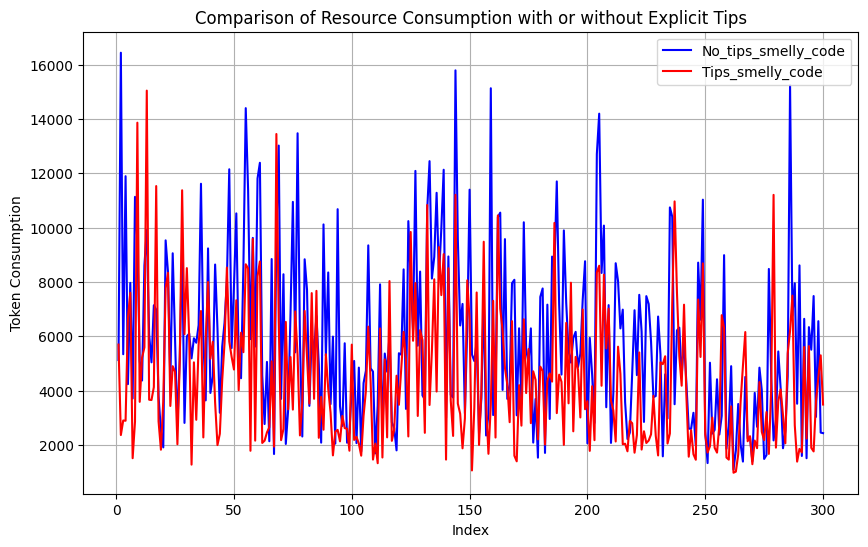} 
  \caption{Comparison of total token consumption for smelly code with and without explicit prompt tips} 
  \label{fig:token_tips_effect} 
\end{figure}

\paragraph{Analysis}
Our study, conducted with a 70\% code functionality similarity threshold (Tab.~\ref{tab:code_similarity_comparison}). During analysis, we examined the token consumption profiles under each prompting condition. Results (Fig.~\ref{fig:token_tips_effect})(Tab.~\ref{tab:token_tips_comparison}) indicate that adding explicit smell annotations to the prompt yields a more efficient reasoning process, with a marked reduction in token usage. Explicit annotations supply the model with rich contextual cues about potential issues, enabling it to bypass redundant reasoning and validation steps. In contrast, without smell hints, the model must engage in multiple verification cycles to detect and interpret hidden code complexities, thereby inflating token consumption.

\begin{table}[htbp]
	\caption{Descriptive Statistics of Total Token Consumption for Smelly Code (With vs. Without Prompt Tips)}
	\begin{center}
		\begin{tabular}{p{3.5cm}|c|c}
			\hline
			\textbf{Statistic} & \textbf{No Tips} & \textbf{With Tips} \\
			\hline
			Count             & 300      & 300      \\
			Mean              & \textbf{5876.49}  & 4431.19  \\
			Std. Dev.         & \textbf{3172.77}  & 2578.35  \\
			Min               & \textbf{1095.00}  & 981.00   \\
			25\% Quartile     & \textbf{3419.50}  & 2312.50  \\
			Median (50\%)     & \textbf{5351.00}  & 3685.00  \\
			75\% Quartile     & \textbf{7985.75}  & 5820.00  \\
			Max               & \textbf{16440.00} & 15045.00 \\
			\hline
		\end{tabular}
		\label{tab:token_tips_comparison}
	\end{center}
\end{table}

\begin{table}[htbp]
\caption{Comparison of CodeBLEU and Similarity Metrics with and without Tips}
\centering
\begin{tabular}{l|c|c}
	\hline
	\textbf{Metric} & \textbf{Mean (With Tips)} & \textbf{Mean (No Tips)} \\
	\hline
	{\textbf{CodeBLEU}} 
	& 0.1585 & 0.1665 \\
	\hline
	{\textbf{Code Similarity}} 
	& 0.7248 & 0.7619 \\
	\hline
\end{tabular}
\label{tab:code_similarity_comparison}
\end{table}

\begin{tcolorbox}[myfindings,title=Findings\,--\,Explicit Smell Prompt]
Explicitly indicating code smell types in the model prompt reduces token consumption by approximately 24.5\% compared to unannotated prompts. The provision of clear smell-specific cues allows the model to focus its reasoning on relevant code fragments, eliminating unnecessary validation steps and lowering computational overhead. This strategy thus significantly enhances inference efficiency and mitigates resource usage in large‑model code reasoning.
\end{tcolorbox}

\subsection{RQ5: Prompt Engineering Strategies beyond Refactoring}

\begin{table*}[htbp]
  \caption{Mean Values of Key Metrics under Different Strategies}
  \centering
  \scriptsize
  \begin{tabular}{ccccc}
    \hline
    \textbf{Strategy}
      & \textbf{Comp‑Norm Code}
      & \textbf{Line‑Scaled Code}
      & \textbf{Comp‑Norm Token}
      & \textbf{Line‑Scaled Token} \\
    \hline
    Base      & 0.1015 & 46.7589  & 0.6026 & 277.7973 \\
    Context   & 0.0858 & 39.3372  & 0.4979 & 229.0921 \\
    Func      & 0.0891 & 40.9848  & 0.5357 & 241.2648 \\
    Total     & 0.0981 & 44.9317  & 0.4760 & 220.8598 \\
    DevOps    & 0.0852 & 38.5346  & 0.4832 & 218.3642 \\
    QAer      & 0.0868 & 40.4725  & 0.5362 & 238.8191 \\
    SEer      & 0.0771 & 35.5723  & 0.5535 & 241.9891 \\
    AbsCost   & 0.0815 & 37.1547  & 0.4746 & 202.5059 \\
    RelCost   & 0.0859 & 38.7631  & 0.4038 & 178.0056 \\
    Comb1     & 0.1001 & 44.5918  & 0.5705 & 245.4126 \\
    Comb2     & 0.0981 & 44.1189  & 0.5301 & 227.1736 \\
    \hline
  \end{tabular}
  \label{tab:transposed_means}
\end{table*}

\begin{table*}[htbp]
  \caption{CodeBLEU and Docstring Similarity Statistics}
  \centering
  \begin{tabular}{l|c|c|c}
  \hline
  \textbf{Category} & \textbf{Method} & \textbf{Mean CodeBLEU} & \textbf{Mean Code-and-GC-Code Similarity} \\
  \hline
  \multirow{1}{*}{Baseline} & - & 0.1803 & 0.7364 \\
  \hline
  \multirow{3}{*}{Context Awareness} & Code Context & 0.1783 & 0.7428 \\
  & Code Function & 0.2283 & 0.7668 \\
  & All Information & 0.2464 & 0.7828 \\
  \hline
  \multirow{3}{*}{Responsibility Tuning} & SEer & 0.1593 & 0.7210 \\
  & QAer & 0.1590 & 0.7369 \\
  & DevOps & 0.1570 & 0.7275 \\
  \hline
  \multirow{2}{*}{Cost Sensitive} & Absolutely & 0.1605 & 0.7011 \\
  & Relatively & 0.1799 & 0.6988 \\
  \hline
  \multirow{2}{*}{Combination} & All Methods & 0.2513 & 0.7635 \\
  & All Methods & 0.2505 & 0.7665 \\
  \hline
  \end{tabular}
  \label{tab:nano-surge-equal}
\end{table*}

\paragraph{Experimental Design / Setup}
To address RQ5, we designed a series of experiments(Fig.~\ref{fig:nano-surge}) aimed at evaluating three prompt engineering strategies—Context Awareness, Responsibility Tuning, and Cost Sensitivity—and their effect on reducing token consumption during the model’s reasoning over smelly code. Each strategy modifies the input prompt to guide the model’s cognitive process, driving a more efficient and focused inference, while minimizing unnecessary token expenditure. Specifically:

\begin{itemize}
  \item \textbf{Context Awareness:} By infusing the prompts with contextual details such as file path, module or function names, and surrounding code snippets, we enhanced the model’s comprehension of the code’s role within the broader system. This enriched context directs the model’s focus to key segments of the code, thereby avoiding redundant inference over irrelevant portions.
  \item \textbf{Responsibility Tuning:} Through the strategic assignment of roles—such as “software engineer,” “QA engineer,” or “DevOps engineer”—we framed the model’s reasoning process to align with a specific perspective. This role-induced focus curtailed extraneous reasoning steps and unnecessary pathways, leading to a more efficient processing of the code.
  \item \textbf{Cost Sensitivity:} We introduced explicit limits on the length of the generated output or the token count (e.g., maximum token budget). By constraining the token size, we controlled the model's computational cost, particularly when generating larger code segments, effectively reducing token expenditure without sacrificing the integrity of the generation.
\end{itemize}

All experiments were performed on the same dataset of 300 smelly code snippets. For each condition, we invoked DeepSeek‑R1, recording the total token consumption and calculating token consumption per unit time. Results were compared against a baseline prompt, which did not include any specific optimization strategies.

\paragraph{Analysis}
We analyzed token consumption across different prompt conditions by calculating tokens per unit time for each strategy. Our key observations include(Tab.~\ref{tab:transposed_means}):

\begin{itemize}
  \item \textbf{Context Awareness:} By enriching the prompts with contextual information, the model was able to bypass irrelevant logic and focus on the core components of the code. On average, token consumption decreased by 15–20\%, with more significant reductions observed in longer, more complex code fragments.
  \item \textbf{Responsibility Tuning:} Assigning roles to the model significantly honed its focus on task‑relevant reasoning steps. This strategy yielded a 10–15\% reduction in token consumption, particularly in large‑scale codebases, where the model’s focus was sharpened by the specified role.
  \item \textbf{Cost Sensitivity:} Imposing limits on output length or token budget led to a 20–30\% reduction in token expenditure. However, overly stringent constraints sometimes led to the truncation of essential code elements, resulting in reduced functional correctness and lower accuracy in the output.
\end{itemize}

\begin{tcolorbox}[myfindings,title=Findings\,--\,Prompt Optimization Strategies]
	Context Awareness and Responsibility Tuning proved most effective for reducing token consumption beyond refactoring (Tab.~\ref{tab:nano-surge-equal}). Context Awareness reduced token usage by 15–20\%, especially for complex code, while Responsibility Tuning achieved a 10–15\% reduction by specifying the model’s reasoning role. Cost Sensitivity reduced tokens by up to 30\% but sometimes caused incomplete outputs. Overall, combining context-rich prompts with clear role framing offered notable efficiency gains without sacrificing functionality.
	
	These optimizations also improved code quality: codeBLEU scores increased by 38\%, and function similarity improved by 4\% compared to the baseline, demonstrating that prompt engineering can simultaneously enhance efficiency and output quality.
\end{tcolorbox}

\section{Discussion}
This study has conducted preliminary observations and analyses of the token consumption behavior exhibited by Large Language Models (LLMs) in Chain of Thought (CoT) scenarios\cite{hou2024largelanguagemodelssoftware}. The findings reveal complex interactions between code characteristics and token overhead, alongside several possibilities for optimization through external strategies\cite{zhu2025chainofthoughttokenscomputerprogram}. The specific behavior of LLMs in software engineering tasks, particularly their resource consumption patterns (such as token usage), often demonstrates a high degree of contextual dependency and certain "black box" characteristics, rendering precise prediction solely through theoretical models challenging.

In this context, the long-standing empirical tradition within software engineering—which emphasizes understanding and mastering complex systems through iteration, observation, experimentation, and data-driven analysis—offers a potent perspective and methodology for addressing such challenges\cite{info16020073}. When confronting the specific issue of token consumption by LLMs in code processing, an approach grounded in empiricism is particularly crucial\cite{fan2025cothinktokenefficientreasoninginstruct}. Empirically observing the actual impact of various factors, such as different code features and prompting strategies, on token consumption, and quantitatively evaluating the effectiveness of intervention measures (like the code refactoring and prompt optimization explored in this study), serves as an effective pathway to deeply understand and manage this novel form of "resource consumption." The analytical framework of the present study represents a preliminary endeavor guided by this philosophy.

Looking ahead, the role of LLMs in software development and associated areas of focus are expected to continually evolve. On one hand, as the capabilities of LLMs iteratively improve, the conformity of the code they generate, or their ability to understand and intelligently refactor existing code\cite{10.5555/311424}, is poised for significant enhancement. This may gradually reduce the future need for manual attention to, and remediation of, traditional "code smells," potentially shifting the focus of software quality assurance to new dimensions. On the other hand, even if LLMs can efficiently process or generate "smell-free," high-quality code, the reasoning and generation processes essential for fulfilling complex requirements will inherently continue to incur token costs\cite{velasco2025propenselargelanguagemodels}. Therefore, regardless of how the issue of "smells" at the code level evolves, attention to token consumption will remain a core concern in the field of LLM-assisted software development. Optimizing the token efficiency of the generation process and striking the optimal balance between output quality, functional completeness, and resource expenditure will continue to be key challenges and top priorities for future research aimed at the sustainable application of this technology\cite{rasnayaka2024empiricalstudyusageperceptions}.

\section{Related Work}

Over recent years, with the widespread application of large language models (LLMs) in natural language processing (NLP)\cite{hou2024largelanguagemodelssoftware}, automated code generation and optimization have also made significant strides\cite{yang2024chainofthoughtneuralcodegeneration}. Research in this area can be broadly categorized into three directions:

\subsection{Code Smell Detection and Refactoring}
Code smells are syntactic or structural code features that do not immediately impair program functionality but degrade readability, maintainability, and extensibility\cite{zhang2022codesmellsmachinelearning}\cite{10.5555/311424}\cite{1357825}. Tufano \textit{et al.} (2017)\cite{mastropaolo2021studyingusagetexttotexttransfer} proposed an automated smell detection approach by analyzing code structures and patterns to identify potential smell instances. Lanza (2003)\cite{1232284} further investigated the impact of code smells on program maintainability and introduced corresponding refactoring strategies. Many refactoring tools combine rule‑based and machine learning techniques to automatically detect and repair smells at scale\cite{zhang2022codesmellsmachinelearning}. Nevertheless, efficiently handling complex smells remains challenging, particularly in deep learning–based inference where token consumption and computational overhead can escalate.

\subsection{LLM‑based Code Generation and Optimization}
The advent of pre‑trained language models has spurred extensive research into LLM‑driven code generation. Models such as OpenAI’s GPT series and Codex, pre‑trained on large code corpora, demonstrate strong capabilities in code completion, synthesis, and repair. The Transformer architecture of Vaswani \textit{et al.} (2017) laid the foundation for these tasks, and subsequent variants such as BERT and GPT have further advanced code generation quality. While LLMs capture syntactic and semantic patterns effectively, their inference often suffers from token inflation, especially when processing complex or smelly code. To mitigate this, researchers have explored integrating automated refactoring with LLM training to jointly optimize code quality and inference efficiency.

\subsection{Integration of Code Quality and Inference Efficiency}
Recently, there has been growing interest in coupling code quality improvements with inference efficiency in pre‑trained models. Pre‑trained models like DeepCode and CodeBERT exhibit strong performance in code quality assessment and automated repair. Empirical studies indicate that higher code quality not only enhances maintainability but also reduces inference costs. For example, Zhang \textit{et al.} (2020) proposed a deep learning–based optimization method that produces functionally equivalent code while significantly lowering token consumption during inference. Building upon this work, our study investigates how code smell removal and prompt engineering jointly reduce token overhead, particularly on complex code fragments.

\section{Conclusion and Future Work}

This study investigated strategies to optimize the reasoning pipeline of large language models (LLMs) in order to reduce token consumption when processing code with smells\cite{zhang2024comprehensiveevaluationparameterefficientfinetuning}\cite{lu2021codexgluemachinelearningbenchmark}. We proposed a combined approach of automated code refactoring and prompt engineering\cite{cordeiro2024empiricalstudycoderefactoring}, and empirically demonstrated that these strategies significantly enhance inference efficiency\cite{guo2024stopefficientcodegeneration}. Our experimental results show that code smells substantially inflate token usage during inference, whereas eliminating smells via refactoring and refining prompts markedly reduces computational overhead\cite{10.5555/311424}\cite{1357825}. Specifically, code refactoring alone decreased token consumption by approximately 50\%, explicit smell annotations in the prompt yielded a 24.5\% reduction, and prompt engineering techniques such as Context Awareness and Responsibility Tuning provided further savings.

Despite these advances, several limitations and avenues for future work remain. First, our experiments focused exclusively on Java and the DeepSeek‑R1 model; future research should extend our methodology to other programming languages and inference engines to assess its generalizability. Second, while refactoring and prompt optimizations proved effective, complex code‐smell scenarios may still challenge inference efficiency\cite{guo2024stopefficientcodegeneration}. Enhancing model reasoning capabilities for diverse programming tasks and more intricate smell types warrants deeper investigation.

Moreover, current work has largely concentrated on token consumption and inference speed\cite{li202512surveyreasoning}. Future studies should explore the trade‐off between generation quality and efficiency—for example, how to maintain code correctness, maintainability, and extensibility while further reducing token usage\cite{liu2024promptlearningmultilabelcode}. Integrating automated refactoring with deep learning–based optimization could also offer practical guidance for improving code quality in large codebases and advancing intelligent programming tools.

In summary, this research provides new insights into token optimization in code generation and reasoning, highlighting the pivotal role of smell elimination and prompt refinement. As intelligent coding assistants evolve, we anticipate that more granular reasoning optimizations and strategic prompt designs will enable next‐generation code generation systems to meet real‐world demands with greater efficiency and precision.

% \section*{References}
% \printbibliography  % 输出参考文献

% Please number citations consecutively within brackets \cite{b1}. The 
% sentence punctuation follows the bracket \cite{b2}. Refer simply to the reference 
% number, as in \cite{b3}---do not use ``Ref. \cite{b3}'' or ``reference \cite{b3}'' except at 
% the beginning of a sentence: ``Reference \cite{b3} was the first $\ldots$''

% Number footnotes separately in superscripts. Place the actual footnote at 
% the bottom of the column in which it was cited. Do not put footnotes in the 
% abstract or reference list. Use letters for table footnotes.

% Unless there are six authors or more give all authors' names; do not use 
% ``et al.''. Papers that have not been published, even if they have been 
% submitted for publication, should be cited as ``unpublished'' \cite{b4}. Papers 
% that have been accepted for publication should be cited as ``in press'' \cite{b5}. 
% Capitalize only the first word in a paper title, except for proper nouns and 
% element symbols.

% For papers published in translation journals, please give the English 
% citation first, followed by the original foreign-language citation \cite{b6}.

\bibliographystyle{IEEEtran}  % IEEE 参考文献格式
\bibliography{references}     % 引用 .bib 文件

\begin{thebibliography}{10}
\providecommand{\url}[1]{#1}
\csname url@samestyle\endcsname
\providecommand{\newblock}{\relax}
\providecommand{\bibinfo}[2]{#2}
\providecommand{\BIBentrySTDinterwordspacing}{\spaceskip=0pt\relax}
\providecommand{\BIBentryALTinterwordstretchfactor}{4}
\providecommand{\BIBentryALTinterwordspacing}{\spaceskip=\fontdimen2\font plus
\BIBentryALTinterwordstretchfactor\fontdimen3\font minus
  \fontdimen4\font\relax}
\providecommand{\BIBforeignlanguage}[2]{{%
\expandafter\ifx\csname l@#1\endcsname\relax
\typeout{** WARNING: IEEEtran.bst: No hyphenation pattern has been}%
\typeout{** loaded for the language `#1'. Using the pattern for}%
\typeout{** the default language instead.}%
\else
\language=\csname l@#1\endcsname
\fi
#2}}
\providecommand{\BIBdecl}{\relax}
\BIBdecl

\bibitem{deepseekai2025deepseekr1incentivizingreasoningcapability}
\BIBentryALTinterwordspacing
DeepSeek-AI, D.~Guo, D.~Yang, H.~Zhang, J.~Song, R.~Zhang, R.~Xu, Q.~Zhu,
  S.~Ma, P.~Wang, X.~Bi, X.~Zhang, X.~Yu, Y.~Wu, Z.~F. Wu, Z.~Gou, Z.~Shao,
  Z.~Li, Z.~Gao, A.~Liu, B.~Xue, B.~Wang, B.~Wu, B.~Feng, C.~Lu, C.~Zhao,
  C.~Deng, C.~Zhang, C.~Ruan, D.~Dai, D.~Chen, D.~Ji, E.~Li, F.~Lin, F.~Dai,
  F.~Luo, G.~Hao, G.~Chen, G.~Li, H.~Zhang, H.~Bao, H.~Xu, H.~Wang, H.~Ding,
  H.~Xin, H.~Gao, H.~Qu, H.~Li, J.~Guo, J.~Li, J.~Wang, J.~Chen, J.~Yuan,
  J.~Qiu, J.~Li, J.~L. Cai, J.~Ni, J.~Liang, J.~Chen, K.~Dong, K.~Hu, K.~Gao,
  K.~Guan, K.~Huang, K.~Yu, L.~Wang, L.~Zhang, L.~Zhao, L.~Wang, L.~Zhang,
  L.~Xu, L.~Xia, M.~Zhang, M.~Zhang, M.~Tang, M.~Li, M.~Wang, M.~Li, N.~Tian,
  P.~Huang, P.~Zhang, Q.~Wang, Q.~Chen, Q.~Du, R.~Ge, R.~Zhang, R.~Pan,
  R.~Wang, R.~J. Chen, R.~L. Jin, R.~Chen, S.~Lu, S.~Zhou, S.~Chen, S.~Ye,
  S.~Wang, S.~Yu, S.~Zhou, S.~Pan, S.~S. Li, S.~Zhou, S.~Wu, S.~Ye, T.~Yun,
  T.~Pei, T.~Sun, T.~Wang, W.~Zeng, W.~Zhao, W.~Liu, W.~Liang, W.~Gao, W.~Yu,
  W.~Zhang, W.~L. Xiao, W.~An, X.~Liu, X.~Wang, X.~Chen, X.~Nie, X.~Cheng,
  X.~Liu, X.~Xie, X.~Liu, X.~Yang, X.~Li, X.~Su, X.~Lin, X.~Q. Li, X.~Jin,
  X.~Shen, X.~Chen, X.~Sun, X.~Wang, X.~Song, X.~Zhou, X.~Wang, X.~Shan, Y.~K.
  Li, Y.~Q. Wang, Y.~X. Wei, Y.~Zhang, Y.~Xu, Y.~Li, Y.~Zhao, Y.~Sun, Y.~Wang,
  Y.~Yu, Y.~Zhang, Y.~Shi, Y.~Xiong, Y.~He, Y.~Piao, Y.~Wang, Y.~Tan, Y.~Ma,
  Y.~Liu, Y.~Guo, Y.~Ou, Y.~Wang, Y.~Gong, Y.~Zou, Y.~He, Y.~Xiong, Y.~Luo,
  Y.~You, Y.~Liu, Y.~Zhou, Y.~X. Zhu, Y.~Xu, Y.~Huang, Y.~Li, Y.~Zheng, Y.~Zhu,
  Y.~Ma, Y.~Tang, Y.~Zha, Y.~Yan, Z.~Z. Ren, Z.~Ren, Z.~Sha, Z.~Fu, Z.~Xu,
  Z.~Xie, Z.~Zhang, Z.~Hao, Z.~Ma, Z.~Yan, Z.~Wu, Z.~Gu, Z.~Zhu, Z.~Liu, Z.~Li,
  Z.~Xie, Z.~Song, Z.~Pan, Z.~Huang, Z.~Xu, Z.~Zhang, and Z.~Zhang,
  ``Deepseek-r1: Incentivizing reasoning capability in llms via reinforcement
  learning,'' 2025. [Online]. Available: \url{https://arxiv.org/abs/2501.12948}
\BIBentrySTDinterwordspacing

\bibitem{cao2023studypromptdesignadvantages}
\BIBentryALTinterwordspacing
J.~Cao, M.~Li, M.~Wen, and S.~chi Cheung, ``A study on prompt design,
  advantages and limitations of chatgpt for deep learning program repair,''
  2023. [Online]. Available: \url{https://arxiv.org/abs/2304.08191}
\BIBentrySTDinterwordspacing

\bibitem{nikiema2025code}
S.~L. Nikiema, J.~Samhi, A.~K. Kabor{\'e}, J.~Klein, and T.~F. Bissyand{\'e},
  ``The code barrier: What llms actually understand?'' \emph{arXiv preprint
  arXiv:2504.10557}, 2025.

\bibitem{hou2024largelanguagemodelssoftware}
\BIBentryALTinterwordspacing
X.~Hou, Y.~Zhao, Y.~Liu, Z.~Yang, K.~Wang, L.~Li, X.~Luo, D.~Lo, J.~Grundy, and
  H.~Wang, ``Large language models for software engineering: A systematic
  literature review,'' 2024. [Online]. Available:
  \url{https://arxiv.org/abs/2308.10620}
\BIBentrySTDinterwordspacing

\bibitem{kang2025distillingllmagentsmall}
\BIBentryALTinterwordspacing
M.~Kang, J.~Jeong, S.~Lee, J.~Cho, and S.~J. Hwang, ``Distilling llm agent into
  small models with retrieval and code tools,'' 2025. [Online]. Available:
  \url{https://arxiv.org/abs/2505.17612}
\BIBentrySTDinterwordspacing

\bibitem{guo2024stopefficientcodegeneration}
\BIBentryALTinterwordspacing
L.~Guo, Y.~Wang, E.~Shi, W.~Zhong, H.~Zhang, J.~Chen, R.~Zhang, Y.~Ma, and
  Z.~Zheng, ``When to stop? towards efficient code generation in llms with
  excess token prevention,'' 2024. [Online]. Available:
  \url{https://arxiv.org/abs/2407.20042}
\BIBentrySTDinterwordspacing

\bibitem{Dolata_2024}
\BIBentryALTinterwordspacing
M.~Dolata, N.~Lange, and G.~Schwabe, ``Development in times of hype: How
  freelancers explore generative ai?'' in \emph{Proceedings of the IEEE/ACM
  46th International Conference on Software Engineering}, ser. ICSE
  ’24.\hskip 1em plus 0.5em minus 0.4em\relax ACM, Apr. 2024, p. 1–13.
  [Online]. Available: \url{http://dx.doi.org/10.1145/3597503.3639111}
\BIBentrySTDinterwordspacing

\bibitem{luo2025autol2sautolongshortreasoning}
\BIBentryALTinterwordspacing
F.~Luo, Y.-N. Chuang, G.~Wang, H.~A.~D. Le, S.~Zhong, H.~Liu, J.~Yuan, Y.~Sui,
  V.~Braverman, V.~Chaudhary, and X.~Hu, ``Autol2s: Auto long-short reasoning
  for efficient large language models,'' 2025. [Online]. Available:
  \url{https://arxiv.org/abs/2505.22662}
\BIBentrySTDinterwordspacing

\bibitem{Alarcia_2024}
\BIBentryALTinterwordspacing
R.~M.~G. Alarcia, ``Optimizing token usage on large language model
  conversations using the design structure matrix,'' in \emph{Proceedings of
  the 26th International DSM Conference (DSM 2024), Stuttgart, Germany}, ser.
  DSM 2024.\hskip 1em plus 0.5em minus 0.4em\relax The Design Society, 2024, p.
  069–078. [Online]. Available: \url{http://dx.doi.org/10.35199/dsm2024.08}
\BIBentrySTDinterwordspacing

\bibitem{yeo2025demystifyinglongchainofthoughtreasoning}
\BIBentryALTinterwordspacing
E.~Yeo, Y.~Tong, M.~Niu, G.~Neubig, and X.~Yue, ``Demystifying long
  chain-of-thought reasoning in llms,'' 2025. [Online]. Available:
  \url{https://arxiv.org/abs/2502.03373}
\BIBentrySTDinterwordspacing

\bibitem{1232284}
M.~Lanza and S.~Ducasse, ``Polymetric views - a lightweight visual approach to
  reverse engineering,'' \emph{IEEE Transactions on Software Engineering},
  vol.~29, no.~9, pp. 782--795, 2003.

\bibitem{10.5555/311424}
\emph{Refactoring: improving the design of existing code}.\hskip 1em plus 0.5em
  minus 0.4em\relax USA: Addison-Wesley Longman Publishing Co., Inc., 1999.

\bibitem{1357825}
M.~Mantyla, J.~Vanhanen, and C.~Lassenius, ``Bad smells - humans as code
  critics,'' in \emph{20th IEEE International Conference on Software
  Maintenance, 2004. Proceedings.}, 2004, pp. 399--408.

\bibitem{zhang2024comprehensiveevaluationparameterefficientfinetuning}
\BIBentryALTinterwordspacing
B.~Zhang, P.~Liang, X.~Zhou, X.~Zhou, D.~Lo, Q.~Feng, Z.~Li, and L.~Li, ``A
  comprehensive evaluation of parameter-efficient fine-tuning on method-level
  code smell detection,'' 2024. [Online]. Available:
  \url{https://arxiv.org/abs/2412.13801}
\BIBentrySTDinterwordspacing

\bibitem{zakharov2025aisoftwareengineeringperceived}
\BIBentryALTinterwordspacing
I.~Zakharov, E.~Koshchenko, and A.~Sergeyuk, ``Ai in software engineering:
  Perceived roles and their impact on adoption,'' 2025. [Online]. Available:
  \url{https://arxiv.org/abs/2504.20329}
\BIBentrySTDinterwordspacing

\bibitem{nghiem2024envisioningnextgenerationaicoding}
\BIBentryALTinterwordspacing
K.~Nghiem, A.~M. Nguyen, and N.~D.~Q. Bui, ``Envisioning the next-generation ai
  coding assistants: Insights \& proposals,'' 2024. [Online]. Available:
  \url{https://arxiv.org/abs/2403.14592}
\BIBentrySTDinterwordspacing

\bibitem{sun2025surveyneuralcodeintelligence}
\BIBentryALTinterwordspacing
Q.~Sun, Z.~Chen, F.~Xu, K.~Cheng, C.~Ma, Z.~Yin, J.~Wang, C.~Han, R.~Zhu,
  S.~Yuan, Q.~Guo, X.~Qiu, P.~Yin, X.~Li, F.~Yuan, L.~Kong, X.~Li, and Z.~Wu,
  ``A survey of neural code intelligence: Paradigms, advances and beyond,''
  2025. [Online]. Available: \url{https://arxiv.org/abs/2403.14734}
\BIBentrySTDinterwordspacing

\bibitem{zhang2024datapreparationdeeplearning}
\BIBentryALTinterwordspacing
F.~Zhang, Z.~Zhang, J.~W. Keung, X.~Tang, Z.~Yang, X.~Yu, and W.~Hu, ``Data
  preparation for deep learning based code smell detection: A systematic
  literature review,'' 2024. [Online]. Available:
  \url{https://arxiv.org/abs/2406.19240}
\BIBentrySTDinterwordspacing

\bibitem{cordeiro2024empiricalstudycoderefactoring}
\BIBentryALTinterwordspacing
J.~Cordeiro, S.~Noei, and Y.~Zou, ``An empirical study on the code refactoring
  capability of large language models,'' 2024. [Online]. Available:
  \url{https://arxiv.org/abs/2411.02320}
\BIBentrySTDinterwordspacing

\bibitem{xie2023uncertaintyawaremoleculardynamicsbayesian}
\BIBentryALTinterwordspacing
Y.~Xie, J.~Vandermause, S.~Ramakers, N.~H. Protik, A.~Johansson, and
  B.~Kozinsky, ``Uncertainty-aware molecular dynamics from bayesian active
  learning for phase transformations and thermal transport in sic,'' 2023.
  [Online]. Available: \url{https://arxiv.org/abs/2203.03824}
\BIBentrySTDinterwordspacing

\bibitem{hasan2025opensourceaipoweredoptimizationscalene}
\BIBentryALTinterwordspacing
S.~Hasan and S.~Basak, ``Open-source ai-powered optimization in scalene:
  Advancing python performance profiling with deepseek-r1 and llama 3.2,''
  2025. [Online]. Available: \url{https://arxiv.org/abs/2502.10299}
\BIBentrySTDinterwordspacing

\bibitem{khojah2024impactpromptprogrammingfunctionlevel}
\BIBentryALTinterwordspacing
R.~Khojah, F.~G. de~Oliveira~Neto, M.~Mohamad, and P.~Leitner, ``The impact of
  prompt programming on function-level code generation,'' 2024. [Online].
  Available: \url{https://arxiv.org/abs/2412.20545}
\BIBentrySTDinterwordspacing

\bibitem{liu2024promptlearningmultilabelcode}
\BIBentryALTinterwordspacing
H.~Liu, Y.~Zhang, V.~Saikrishna, Q.~Tian, and K.~Zheng, ``Prompt learning for
  multi-label code smell detection: A promising approach,'' 2024. [Online].
  Available: \url{https://arxiv.org/abs/2402.10398}
\BIBentrySTDinterwordspacing

\bibitem{zhu2025chainofthoughttokenscomputerprogram}
\BIBentryALTinterwordspacing
F.~Zhu, P.~Wang, and Z.~Sui, ``Chain-of-thought tokens are computer program
  variables,'' 2025. [Online]. Available:
  \url{https://arxiv.org/abs/2505.04955}
\BIBentrySTDinterwordspacing

\bibitem{info16020073}
\BIBentryALTinterwordspacing
Z.~Li and X.~Lu, ``Research on compressed input sequences based on compiler
  tokenization,'' \emph{Information}, vol.~16, no.~2, 2025. [Online].
  Available: \url{https://www.mdpi.com/2078-2489/16/2/73}
\BIBentrySTDinterwordspacing

\bibitem{fan2025cothinktokenefficientreasoninginstruct}
\BIBentryALTinterwordspacing
S.~Fan, P.~Han, S.~Shang, Y.~Wang, and A.~Sun, ``Cothink: Token-efficient
  reasoning via instruct models guiding reasoning models,'' 2025. [Online].
  Available: \url{https://arxiv.org/abs/2505.22017}
\BIBentrySTDinterwordspacing

\bibitem{velasco2025propenselargelanguagemodels}
\BIBentryALTinterwordspacing
A.~Velasco, D.~Rodriguez-Cardenas, L.~R. Alif, D.~N. Palacio, and
  D.~Poshyvanyk, ``How propense are large language models at producing code
  smells? a benchmarking study,'' 2025. [Online]. Available:
  \url{https://arxiv.org/abs/2412.18989}
\BIBentrySTDinterwordspacing

\bibitem{rasnayaka2024empiricalstudyusageperceptions}
\BIBentryALTinterwordspacing
S.~Rasnayaka, G.~Wang, R.~Shariffdeen, and G.~N. Iyer, ``An empirical study on
  usage and perceptions of llms in a software engineering project,'' 2024.
  [Online]. Available: \url{https://arxiv.org/abs/2401.16186}
\BIBentrySTDinterwordspacing

\bibitem{yang2024chainofthoughtneuralcodegeneration}
\BIBentryALTinterwordspacing
G.~Yang, Y.~Zhou, X.~Chen, X.~Zhang, T.~Y. Zhuo, and T.~Chen,
  ``Chain-of-thought in neural code generation: From and for lightweight
  language models,'' 2024. [Online]. Available:
  \url{https://arxiv.org/abs/2312.05562}
\BIBentrySTDinterwordspacing

\bibitem{zhang2022codesmellsmachinelearning}
\BIBentryALTinterwordspacing
H.~Zhang, L.~Cruz, and A.~van Deursen, ``Code smells for machine learning
  applications,'' 2022. [Online]. Available:
  \url{https://arxiv.org/abs/2203.13746}
\BIBentrySTDinterwordspacing

\bibitem{mastropaolo2021studyingusagetexttotexttransfer}
\BIBentryALTinterwordspacing
A.~Mastropaolo, S.~Scalabrino, N.~Cooper, D.~N. Palacio, D.~Poshyvanyk,
  R.~Oliveto, and G.~Bavota, ``Studying the usage of text-to-text transfer
  transformer to support code-related tasks,'' 2021. [Online]. Available:
  \url{https://arxiv.org/abs/2102.02017}
\BIBentrySTDinterwordspacing

\bibitem{lu2021codexgluemachinelearningbenchmark}
\BIBentryALTinterwordspacing
S.~Lu, D.~Guo, S.~Ren, J.~Huang, A.~Svyatkovskiy, A.~Blanco, C.~Clement,
  D.~Drain, D.~Jiang, D.~Tang, G.~Li, L.~Zhou, L.~Shou, L.~Zhou, M.~Tufano,
  M.~Gong, M.~Zhou, N.~Duan, N.~Sundaresan, S.~K. Deng, S.~Fu, and S.~Liu,
  ``Codexglue: A machine learning benchmark dataset for code understanding and
  generation,'' 2021. [Online]. Available:
  \url{https://arxiv.org/abs/2102.04664}
\BIBentrySTDinterwordspacing

\bibitem{li202512surveyreasoning}
\BIBentryALTinterwordspacing
Z.-Z. Li, D.~Zhang, M.-L. Zhang, J.~Zhang, Z.~Liu, Y.~Yao, H.~Xu, J.~Zheng,
  P.-J. Wang, X.~Chen, Y.~Zhang, F.~Yin, J.~Dong, Z.~Guo, L.~Song, and C.-L.
  Liu, ``From system 1 to system 2: A survey of reasoning large language
  models,'' 2025. [Online]. Available: \url{https://arxiv.org/abs/2502.17419}
\BIBentrySTDinterwordspacing

\end{thebibliography}
% \begin{thebibliography}
%   \bibitem{b1}
% \end{thebibliography}

% \bibitem{b1} G. Eason, B. Noble, and I. N. Sneddon, ``On certain integrals of Lipschitz-Hankel type involving products of Bessel functions,'' Phil. Trans. Roy. Soc. London, vol. A247, pp. 529--551, April 1955.
% \bibitem{b2} J. Clerk Maxwell, A Treatise on Electricity and Magnetism, 3rd ed., vol. 2. Oxford: Clarendon, 1892, pp.68--73.
% \bibitem{b3} I. S. Jacobs and C. P. Bean, ``Fine particles, thin films and exchange anisotropy,'' in Magnetism, vol. III, G. T. Rado and H. Suhl, Eds. New York: Academic, 1963, pp. 271--350.
% \bibitem{b4} K. Elissa, ``Title of paper if known,'' unpublished.
% \bibitem{b5} R. Nicole, ``Title of paper with only first word capitalized,'' J. Name Stand. Abbrev., in press.
% \bibitem{b6} Y. Yorozu, M. Hirano, K. Oka, and Y. Tagawa, ``Electron spectroscopy studies on magneto-optical media and plastic substrate interface,'' IEEE Transl. J. Magn. Japan, vol. 2, pp. 740--741, August 1987 [Digests 9th Annual Conf. Magnetics Japan, p. 301, 1982].
% \bibitem{b7} M. Young, The Technical Writer's Handbook. Mill Valley, CA: University Science, 1989.
% \end{thebibliography}
\vspace{12pt}

\end{document}